# Photon allocation strategy in region-of-interest tomographic imaging

Zheyuan Zhu , Hsin-Hisung Huang, and Shuo Pang

*Abstract*— **Photon counting detection is a promising approach toward effectively reducing the radiation dose in x-ray computed tomography (CT). Full CT reconstruction from a fraction of the detected photons required by scintillation-based detectors has been demonstrated. Current efforts in photon-counting CT have focused mainly on reconstruction techniques. In medical and industrial x-ray computed tomography (CT) applications, truncated projection from the region-of-interest (ROI) is another effective way of dose reduction, as information from the ROI is usually sufficient for diagnostic purpose. Projection truncation poses an ill-conditioned inverse problem, which can be improved by including projections from the exterior region. However, this trade-off between the interior reconstruction quality and the additional exterior measurement (extra dose) has not been studied. In this manuscript, we explore the number of detected x-ray photons as a new dimension for measurement engineering. Specifically, we design a flexible, photon-efficient measurement strategy for ROI reconstruction by incorporating the photon statistics at extremely low flux level (~10 photons per pixel). The optimized photon-allocation strategy shows 10 ~ 15-fold lower ROI reconstruction error than truncated projections, and 2-fold lower than whole-volume CT scan. Our analysis in few-photon interior tomography could serve as a new framework for dose-efficient, task-specific x-ray image acquisition design.**

## I. INTRODUCTION

Due to the ionization nature of x-ray radiation, dose reduction is a critical design consideration in x-ray based imaging modality, especially in computed tomography (CT) system where a series of projections are acquired during the acquisition process [1], [2]. Photon counting detectors (PCD) have demonstrated ~30% dose reduction while maintaining the image quality, due to its high detection efficiency [3], [4]. In addition, PCDs can effectively reduce the dark noise and discriminate the unwanted signal through energy gating [5], [6]. In conventional CT scanner, the detector typically collects tens of thousands photons per pixel on average [7]. Image recovery from low-dose measurements necessitate the consideration of photon statistics in the reconstruction [8]. For detected photon counts on the order of $10^2$~$10^3$, Poisson likelihood can be used to model the detected photon counts and infer the attenuation map [9]–[11]. More recently, image reconstruction from an average detection counts of ~10 photons per pixel based on the binomial likelihood [12] has been demonstrated in the visible regime. Extending few-photon imaging framework to CT modality could potentially minimize radiation damage to the object of interest.

In medical or industrial imaging applications, the ultimate task of imaging is often diagnosis or detection of a specific feature, rather than whole-volume reconstruction. Majority of the medical diagnostic scans require high image quality only in a small volume, while the projections outside this volume only provide structural or orientation information [13]. In industrial CT inspection, the whole volume scan could be challenging to manage, yet the image processing tasks focus on classifying or quantifying the localized defects (e.g. air cavity, porosity, etc.) [14]. Task-specific image acquisition design, aiming to measure only the information relevant to the task, can potentially shorten imaging time and reduce radiation damage to the specimen. This paper investigates the acquisition strategies from the perspective of photon statistics, when only a small localized region within the object is of interest.

The image acquisition process designed specifically for ROI reconstruction, termed interior tomography, distributes the radiation exclusively to the ROI, resulting in a series of truncated projections [15]. A unique and stable ROI solution from truncated projections is possible, provided that either a sub-region within the ROI is known [16], [17], or the sample is piecewise constant [18], [19]. However, the additional information of samples may not always be available [20], [21]. Another solution is to use low-resolution projection from the exterior region to stabilize the reconstruction [13], [22]. This approach can be considered as a trade-off between reconstruction stability of whole CT scan and dose reduction benefit of truncated scan. Yet this trade-off has not been quantitatively studied, mainly because the illumination or integration time of each pencil beam is not easily adjustable in a conventional setup [23]. Recently, a time-stamp photon-counting X-ray imaging method has been developed [24]. Instead of counting the number of photons within a fixed integration time, the elapsed time is recorded when a pre-allocated photon count has been reached. With this photon-counting setup, we explore the photon allocation strategies for the localized CT reconstruction, given a fixed total detected photon budget. Based on the statistics of the photon arrival

This work was supported by National Science Foundation under Grant DMS-1615124.

Z. Zhu and S. Pang are with CREOL, The College of Optics and Photonics, University of Central Florida, Orlando, FL, 32816 (e-mails: zyzhu@knights.ucf.edu, pang@creol.ucf.edu)

H.-H Huang is with Department of Statistics, University of Central Florida, Orlando, FL, 32816 (e-mail: Hsin-Hsiung.Huang@ucf.edu)

time, we formulate the tomography reconstruction as a Bayesian estimator. The optimal photon allocation strategy is identified by the least mean square error (MSE) from the interior region, for a fixed total photon budget. ROI reconstructions from an average of ~10 photons per pencil beam have been established in both simulations and experiments. In terms of the ROI reconstruction MSE, the optimized allocation strategy has demonstrated as much as a ~15-fold improvement compared to truncated projection measurements, and ~2-fold compared to a uniform whole volume CT scan with the same total photon budget.

This paper is organized as follows: Section 2 describes the setup and configuration of our photon-counting CT system, and models the measurement and reconstruction process with a negative binomial distribution. Section 3 describes the details of the simulation and experiment setups used to explore the optimal photon allocation strategy. Section 4 presents the ROI reconstruction in phantom simulations, and applies the optimized photon allocation strategy to scan a real sample. Section 5 elaborates on the choice of regularization parameter in the reconstruction, and discusses the difference between our analytical model and the numerical estimator. Section 6 concludes the whole paper.

## II. THEORY

### A. Imaging principle

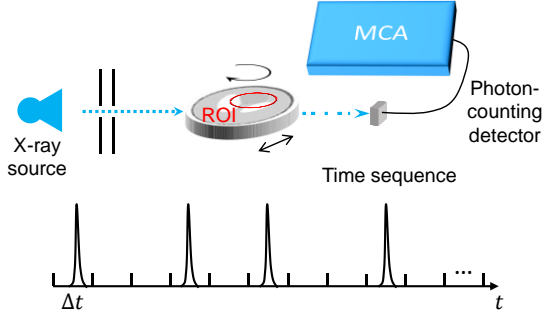

Fig. 1: Illustration of the photon-counting tomography scheme based on the time-stamp of each detected photons. The imaging task is to design an optimized photon allocation strategy among all the measurements to minimize the reconstruction error in ROI, marked by the red circle.

An x-ray tomography system with pencil-beam illumination and a single-pixel photon-counting detector has been constructed in our study. Fig. 1 illustrates the concept of the imaging setup. The incident x-ray beam is collimated by a pair of pinholes. To eliminate dark noise and obtain photon-limited signal, we specify a narrow bandwidth around the Kα line of the source for pulse censoring, so that photons whose energy fall outside the range are rejected. The source flux is well-controlled such that the probability $\lambda$ of detecting one photon within one time interval is much smaller than 1 ($\lambda \sim 10^{-2}$), even when no sample is placed between the source and detector. This ensures that no two photon events could overlap within one time interval. The output signal is a series of discrete time intervals $\Delta t$, within which either one or no photon is registered.

Instead of directly counting the number of photons in a pre-defined integration time, we record the number of elapsed time intervals, $\mathbf{g}$, before the $\mathbf{r}$-th photon is detected for each pencil-beam measurement. To reconstruct the image of a 2D layer, the sample is translated across the pencil beam by $s$, which is the offset between the rotation center and the incident x-ray beam, and rotated around the vertical axis by angle $\phi$.

Let $\mathbf{f} = \{f_{i_x,i_y}\} \in \mathbb{R}_+^{n_x \times n_y}$ denote the 2D attenuation map, where $(i_x, i_y) = \{1, 2 \dots n_x\} \times \{1, 2 \dots n_y\}$ is the pixel index of an $n_x \times n_y$ image. The discrete measurements $\mathbf{g} = \{g_{j_s,j_\phi}\} \in \mathbb{N}^{m_s \times m_\phi}$ and the photons received at each pencil beam $\mathbf{r} = \{r_{j_s,j_\phi}\} \in \mathbb{N}^{m_s \times m_\phi}$ are both indexed by $(j_s, j_\phi) = \{1, 2 \dots m_s\} \times \{1, 2 \dots m_\phi\}$, where $m_s$, $m_\phi$ represent the number of sampling in the translation $s$ and rotation $\phi$ dimension, respectively. In our photon-counting framework, we assign a pre-defined photon number $\mathbf{r}$ to accumulate at each pencil beam, and measure the elapsed time intervals $\mathbf{g}$, which is a negative-binomial random variable. The reconstruction estimates the attenuation map $\hat{\mathbf{f}}$ from the observations of $\mathbf{g}$ with a set of known parameters $\mathbf{r}$. For a given parameter set $\{\mathbf{r}, \mathbf{f}\}$, the ROI reconstruction error is modeled by the weighted mean-square error of the estimator $\hat{\mathbf{f}}$, which is defined as

$$MSE_{\hat{\mathbf{f}}}(\mathbf{r}; \mathbf{f}) = E_{\mathbf{f}} \left\| \mathbf{w} \odot (\hat{\mathbf{f}} - \mathbf{f}) \right\|^2, \quad (1)$$

where $\mathbf{w} = \{w_{i_x,i_y} \in \{0,1\}, (i_x, i_y) = \{1, 2, \dots, n_x\} \times \{1, 2, \dots, n_y\}\}$ denotes the weight of each object pixel; $w_{i_x,i_y}=0$ for pixels outside the ROI, 1 for pixels inside the ROI; $\odot$ denotes the element-wise product between two vectors. The optimal photon allocation strategy looks for a photon map $\mathbf{r}$ that minimizes the ROI reconstruction MSE, given a fixed total detected photon budget

$$\sum_{j_s,j_\phi=1}^{m_s,m_\phi} r_{j_s,j_\phi} = I_0. \quad (2)$$

In the following discussions involving linear indexing, the 2D map $\mathbf{f}$ and $\mathbf{w}$ are vectorized to $n = n_x \times n_y$ vectors $\mathbf{f} = \{f_i, i = 1,2 \dots, n\}$, $\mathbf{w} = \{w_i, i = 1,2 \dots, n\}$; $\mathbf{g}$ and $\mathbf{r}$ to $m = m_s \times m_\phi$ vectors $\mathbf{g} = \{g_j, j = 1,2, \dots, m\}$, $\mathbf{r} = \{r_j, j = 1,2, \dots, m\}$, respectively.

### B. Negative binomial noise model

Considering the sample attenuation, the probability of receiving one photon within one time interval $\Delta t$ for the pencil-beam $j$ is

$$T_j = \lambda \exp(-\sum_{i=1}^n A_{ji} f_i), \quad (3)$$

where $\mathbf{A} \in \mathbf{R}^{m \times n}, A_{ji} \geq 0$ is the CT transform matrix constructed from the distance-driven method [25]. The probability of receiving the $r_j$-th photon at $g_j$-th interval ($g_j, r_j \in N_+, g_j > r_j$) for every pencil beam $j$ follows a negative binomial distribution ($\sim NB(r_j, T_j)$) with explicit parameters $\mathbf{r} = (r_1, r_2, \ldots r_m)$ and implicit parameters $\mathbf{f} = (f_1, f_2, \ldots f_n)$

$$p(\mathbf{g}|\mathbf{f}; \mathbf{r}) = \prod_{j=1}^{m} \binom{g_j - 1}{r_j - 1}(1 - T_j)^{g_j - r_j} T_j^{r_j}. \quad (4)$$

Let $\pi(\mathbf{f})$ denote the prior distribution that describes the smoothness or sparsity constraints on the object. Since the attenuation map is non-negative $\mathbf{f} \geq 0$, we restrain the domain of $\pi(\mathbf{f})$ to non-negative values $\mathbf{f} \in \mathbb{R}_+^n$. The negative log-posterior distribution $L(\mathbf{f}|\mathbf{g}; \mathbf{r})$ of waiting $\mathbf{g}$ intervals for $\mathbf{r}$ photons given the sample prior $\pi(\mathbf{f})$ is

$$\begin{aligned} L(\mathbf{f}|\mathbf{g}; \mathbf{r}) &= -\log\left[\frac{p(\mathbf{g}|\mathbf{f}; \mathbf{r})\pi(\mathbf{f})}{p(\mathbf{g}; \mathbf{r})}\right] \\ &= -\sum_{j=1}^{m}\left\{\log\binom{g_j - 1}{r_j - 1}\right. \\ &\quad + (g_j - r_j)\log\left[1 - \lambda \exp\left(-\sum_{i=1}^{n} A_{ji}f_i\right)\right] \\ &\quad \left. - r_j\sum_{i=1}^{n} A_{ji}f_i\right\} + u(\mathbf{f}) + \log p(\mathbf{g}; \mathbf{r}), \end{aligned} \quad (5)$$

where we introduce $u(\mathbf{f}) = -\log \pi(\mathbf{f})$ as the negative log-prior, $l(\mathbf{f}|\mathbf{g}; \mathbf{r}) = \log p(\mathbf{g}|\mathbf{f}; \mathbf{r})$ as the log-likelihood of the model parameter $\mathbf{f}$ in the negative binomial distribution, for simplicity. The object $\mathbf{f}$ is reconstructed by the maximum-a-posteriori (MAP) estimation, which minimizes the negative log-posterior distribution. The marginal distribution of the measurement $\log p(\mathbf{g}; \mathbf{r})$ is independent of $\mathbf{f}$, and thus not included in the optimization.

$$\hat{\mathbf{f}}(\mathbf{g}; \mathbf{r}) = \underset{\mathbf{f}'}{\mathrm{argmin}}\{L(\mathbf{f}'|\mathbf{g}; \mathbf{r}) = -l(\mathbf{f}'|\mathbf{g}; \mathbf{r}) + u(\mathbf{f}')\}. \quad (6)$$

### C. Property of the measurement and estimator

In this section, we will introduce some quadratic approximations that are essential to analyze the behavior of the numerical MAP estimator (Eq. (6)), for a given object and photon counts $\{\mathbf{r}, \mathbf{f}\}$, and the prior distribution $\pi(\mathbf{f})$. To facilitate the discussion of the measurement distribution, given the parameter set $\{\mathbf{r}, \mathbf{f}\}$, we introduce auxiliary variables

$$\mathbf{t} = \left\{t_j(\mathbf{f}) = \log\frac{\lambda}{T_j(\mathbf{f})} = \sum_{i=1}^{n} A_{ji}f_i, j = 1,2,\ldots,m\right\}, \quad (7)$$

which can be interpreted as the CT line-integral of the object. By setting the derivative of the negative binomial measurement (Eq. (4)) with respect to $T_j$ to 0, and applying the invariance principle, the maximum likelihood (ML) estimation for $t_j$ is achieved at

$$\hat{t}_j = \log\frac{\lambda r_j}{g_j}. \quad (8)$$

We then change the variable in the negative binomial distribution (Eq. (4)) from $\mathbf{g}$ to $\hat{\mathbf{t}}$ to derive the conditional distribution of $\hat{\mathbf{t}}$, $q(\hat{\mathbf{t}}|\mathbf{t}(\mathbf{f}); \mathbf{r})$. The logarithm of $q(\hat{\mathbf{t}}; \mathbf{r})$ is

$$\begin{aligned} q(\hat{\mathbf{t}}|\mathbf{t}; \mathbf{r}) &= \sum_{j=1}^{m} q_j(\hat{t}_j|t_j; r_j) \\ &= \sum_{j=1}^{m}\left\{\log\frac{\Gamma(r_j \exp \hat{t}_j/\lambda)}{\Gamma(r_j \exp \hat{t}_j/\lambda - r_j + 1)\Gamma(r_j)}\right. \\ &\quad + \left(\frac{r_j \exp \hat{t}_j}{\lambda} - r_j\right)\log\left[1 - \lambda \exp(-t_j)\right] \\ &\quad \left. + r_j \log(\lambda \exp(-t_j)) + \log(r_j \exp \hat{t}_j/\lambda)\right\}, \end{aligned} \quad (9)$$

where $\sum_{i=1}^{n} A_{ji}f_i$ is replaced with $t_j$ , $g_j$ with $r_j \exp \hat{t}_j / \lambda$ ; and binomial coefficients in $p(\mathbf{g}|\mathbf{f}; \mathbf{r})$ are expressed by the gamma functions. The additional term $\log(r \exp \hat{t}/\lambda)$ normalizes the distribution with respect to $\hat{\mathbf{t}}$. Based on Eq. (9), we introduce a couple of first-order approximations to the log-likelihood function to simplify the discussion of the distribution of $\hat{t}_j$. Since $\lambda \sim 10^{-2}$, we take the first-order approximation for $\log(1 - \lambda \exp(-t_j)) \approx -\lambda \exp(-t_j)$, and according to Stirling's formula,

$$\frac{\Gamma\left(\frac{r_j \exp \hat{t}_j}{\lambda}\right)}{\Gamma\left(\frac{r_j \exp \hat{t}_j}{\lambda} - r_j + 1\right)} \approx \left(\frac{r_j \exp \hat{t}_j}{\lambda}\right)^{(r_j - 1)},$$

Eq. (9) is simplified to:

$$\begin{aligned} q(\hat{\mathbf{t}}|\mathbf{t}; \mathbf{r}) &= \sum_{j=1}^{m}[r_j(\hat{t}_j - t_j) - r_j \exp(\hat{t}_j - t_j) \\ &\quad + r_j\lambda \exp(-t_j) + r_j \log r_j - \log \Gamma(r_j)]. \end{aligned} \quad (10)$$

Eq. (10) shows that the distribution of $\hat{t}_j$ is asymmetrically centered at $\hat{t}_j = t_j$. Taylor expansion on $q_j(\hat{t}_j|t_j; r_j)$ to the second order around $\hat{t}_j = t_j$ gives

$$\begin{aligned} q_j(\hat{t}_j|t_j; r_j) &= q_j(t_j) + q_j'(t_j)(\hat{t}_j - t_j) \\ &\quad + \frac{q_j''(t_j)(\hat{t}_j - t_j)^2}{2} + R(\hat{t}_j) \\ &= r_j\lambda \exp(-t_j) + r_j \log r_j - \log \Gamma(r_j) \\ &\quad + \frac{r_j(\hat{t}_j - t_j)^2}{2} + R(\hat{t}_j). \end{aligned} \quad (11)$$

The quadratic component in Eq. (11) implies that each random variable $\hat{t}_j$ approximately follows a normal distribution $N(\mu_{\hat{t}_j}, \sigma_{\hat{t}_j}^2)$ with mean $\mu_{\hat{t}_j}$ equal to the ground truth of CT line integral $t_j$, and a variance $\sigma_{\hat{t}_j}^2 = 1/r_j$ that depends only on the photon count. $R(\hat{t}_j)$ contains the higher-order Taylor expansion terms that introduce a negative skew to this normal distribution. When the photon number $r_j$ reaches 16 and above, however, the skew becomes insignificant, which is discussed in Section 5A in detail.

Next, we discuss the property of the MAP estimator with Gaussian prior $\pi(\mathbf{f})$. Here, we omit the non-negativity constraint, and consider the Gaussian family of distributions for obtaining closed-form solutions to the optimization problem in Eq. (6)

$$\pi(\mathbf{f}) = \frac{\sqrt{\tau \det(\mathbf{DD}^T)/2}}{\sqrt{(2\pi)^n}} \exp(-\tau \|\mathbf{Df}\|^2) \qquad (12)$$

where the matrix $\mathbf{D} \in \mathbb{R}^{q \times n}$ projects the attenuation map $\mathbf{f}$ onto a sparse domain; $q$ denotes the number of sparse metrics expressed by the rows of $\mathbf{D}$; the parameter $\tau$ controls the variance of the Gaussian prior. If $\mathbf{D}$ is the identity matrix $\mathbf{I}_n$, $\pi(\mathbf{f})$ becomes the L2-prior that punishes large values in $\mathbf{f}$. After a set of measurements, $\mathbf{g}$, are recorded with photon counts $\mathbf{r}$, we rewrite the likelihood of the parameter $\mathbf{f}$ in terms of the CT line-integral, $\mathbf{t}$, given $\{\mathbf{g}, \mathbf{r}\}$

$$l(\mathbf{f}|\mathbf{g}; \mathbf{r}) = \sum_{j=1}^{m} y_j(t_j)$$
$$= \sum_{j=1}^{m} \left\{ \log \binom{g_j - 1}{r_j - 1} \right. \qquad (13)$$
$$\left. + (g_j - r_j) \log[1 - \lambda \exp(-t_j)] - r_j t_j \right\}.$$

Applying second-order Taylor-expansion to $y_j(t_j)$ around the estimated line-integral $\hat{t}_j = \log(\lambda g_j / r_j)$ yields [26]

$$y_j(t_j) \approx y_j(\hat{t}_j) + y_j'(\hat{t}_j)(t_j - \hat{t}_j) + \frac{y_j''(\hat{t}_j)(t_j - \hat{t}_j)^2}{2}$$
$$= \left\{ \log \binom{g_j - 1}{r_j - 1} + (g_j - r_j) \log[1 - \lambda \exp(-\hat{t}_j)] \right. \qquad (14)$$
$$\left. - r_j \hat{t}_j \right\} - \frac{r_j(t_j - \hat{t}_j)^2}{2\left(1 - \frac{r_j}{g_j}\right)},$$

where the first-order derivative $y_j'(\hat{t}_j)=0$ for all $j = 1,2,\dots,m$; $r_j/g_j = \lambda \exp \sum_{i=1}^{n} A_{ij} f_i \ll 1$ is negligible on the denominator. The Taylor-expansion in Eq. (14) reduces the MAP estimator in Eq. (6) to a least-square problem weighted on photon count $\mathbf{r}$

$$\hat{\mathbf{f}}(\hat{\mathbf{t}}; \mathbf{r}) = \operatorname*{argmin}_{\mathbf{f}'} \left\{ -\sum_{j=1}^{m} y_j(t_j(\mathbf{f}')) + \tau \|\mathbf{Df}'\|_2^2 \right\}$$
$$\approx \operatorname*{argmin}_{\mathbf{f}'} \left\{ \frac{1}{2} \|diag(\mathbf{r})(\mathbf{Af}' - \hat{\mathbf{t}})\|_2^2 + \tau \|\mathbf{Df}'\|_2^2 \right\}, \qquad (15)$$

where the zero-order terms in $y_j$ are independent of $\mathbf{f}$, and can thus be neglected in the optimization; $diag(\mathbf{r})$ denotes the diagonal matrix constructed from the vector $\mathbf{r}$. The resulting objective function in Eq. (15) $\varepsilon(\mathbf{f}) = \frac{1}{2} \|diag(\mathbf{r})(\mathbf{Af} - \hat{\mathbf{t}})\|_2^2 + \tau \|\mathbf{Df}\|_2^2$ has a gradient

$$\nabla \varepsilon(\mathbf{f}) = \mathbf{A}^T diag(\mathbf{r})(\mathbf{Af} - \hat{\mathbf{t}}) + 2\tau \mathbf{D}^T \mathbf{Df} \qquad (16)$$

and Hessian matrix

$$\mathbf{H}(\varepsilon(\mathbf{f})) = \mathbf{A}^T diag(\mathbf{r})\mathbf{A} + 2\tau \mathbf{D}^T \mathbf{D}. \qquad (17)$$

The Hessian matrix is positive-definite for $\mathbf{f} \in \mathbb{R}^n$; therefore the objective function after the approximation (Eq. (15)) is minimized at zero gradient, which has an analytical solution

similar to that of Tikhonov regularization [27]

$$\hat{\mathbf{f}}(\hat{\mathbf{t}}; \mathbf{r}) = [(\mathbf{A}^T diag(\mathbf{r})\mathbf{A} + 2\tau \mathbf{D}^T \mathbf{D})]^{-1} \mathbf{A}^T diag(\mathbf{r}) \hat{\mathbf{t}}. \qquad (18)$$

where $[\cdot]^{-1}$ denotes the pseudo-inverse of the matrix. Eq. (18) shows that each pencil beam measurement, which corresponds to a specific row entry in $\mathbf{A}$, carries a weight according to its photon count, $\mathbf{r}$. A beam receiving higher photon count is assigned with a greater weight because of the smaller uncertainty in the estimated CT line integral. Since the estimator $\hat{\mathbf{f}}$ is a linear superposition of Gaussian variables $\hat{\mathbf{t}}$, the distribution of $\hat{\mathbf{f}}$ thus follows a normal distribution with mean

$$\boldsymbol{\mu}_{\hat{\mathbf{f}}}(\mathbf{r}; \mathbf{f}) = [(\mathbf{A}^T diag(\mathbf{r})\mathbf{A} + 2\tau \mathbf{D}^T \mathbf{D})]^{-1} \mathbf{A}^T diag(\mathbf{r}) \mathbf{Af} \qquad (19)$$

and variance

$$\boldsymbol{\sigma}_{\hat{\mathbf{f}}}^2(\mathbf{r}) = \{[(\mathbf{A}^T diag(\mathbf{r})\mathbf{A} + 2\tau \mathbf{D}^T \mathbf{D})]^{-1} \mathbf{A}^T diag(\mathbf{r})\}^2 \frac{1}{\mathbf{r}}. \qquad (20)$$

The MSE of the estimator comprises the bias square and the variance of the pixels within the ROI

$$MSE_{\hat{\mathbf{f}}}(\mathbf{r}; \mathbf{f}) = \mathbf{w}^T [(\boldsymbol{\mu}_{\hat{\mathbf{f}}}(\mathbf{r}; \mathbf{f}) - \mathbf{f})^2 + \boldsymbol{\sigma}_{\hat{\mathbf{f}}}^2(\mathbf{r})]. \qquad (21)$$

Eq. (19) indicates that without regularization, $\tau=0$, the estimator (Eq. (18)) is unbiased. When $\tau \neq 0$, the Gaussian prior introduces a bias to the estimator and reduces its variance. The tradeoff between bias and variance implies the existence of a best regularization parameter, $\tau$, that yields minimum MSE in the reconstruction.

## III. Materials and Methods

### A. Phantoms and photon maps

Two simulation phantoms were used to evaluate the performance of the estimator (Eq. (6)) with different photon allocation strategies. Simulation phantom 1 was a real abdomen CT slice (Subject ID 116-HM10395) [28]. The raw image underwent 4 X 4 binning to reduce the number of pixels to 80 X 80, with a pixel size of 4mm X 4mm after binning. The simulated pencil beam measurement had a translation step size of 4 mm and a rotation step of $2°$ to cover $180°$ projections. The region around the cross-section of the vertebra was selected as the ROI, which was centered at 80mm (20 pixels) away from the rotation center, and was 80mm in diameter (20-pixel wide). Phantom 2 was a 64 X 64 Shepp-Logan phantom with a pixel size of 1mm. The translation step of the pencil beam was 1mm, matching the spatial grid of the phantom. The rotation covered 0 to $180°$ projections at a step size of $2°$. The 15mm-diameter region (15-pixel wide) was defined as the ROI, which was 4mm (4 pixels) away from the rotation center.

The photon allocation strategy in the ROI and exterior region is modeled by assigning different number of photon counts $\mathbf{r}$ to accumulate at each measurement $j$. To avoid enumerating all possibilities of $\mathbf{r}$ that satisfies the total photon budget constraint, we confine our choice of the 2D photon allocation map to a trapezoid function:

$$
r_{j_s, j_\phi} = \frac{I_0(1-\beta)}{m_\phi} + \begin{cases} \dfrac{1}{m_\phi} \dfrac{I_0 \beta \delta x}{2\sigma + \Delta}, & |j_s \delta x - s_c| \le \sigma \\[3mm] \dfrac{1}{m_\phi} \dfrac{I_0 \beta \delta x}{2\sigma + \Delta} \dfrac{\sigma + \Delta - |j_s \delta x - s_c|}{\Delta}, & \sigma < |j_s \delta x - s_c| < (\sigma + \Delta) \\[3mm] 0, & |j_s \delta x - s_c| \ge (\sigma + \Delta) \end{cases}
\tag{22}
$$

$$
j_s = 1, 2, \ldots, m_s
$$
$$
j_\phi = 1, 2, \ldots, m_\phi
$$

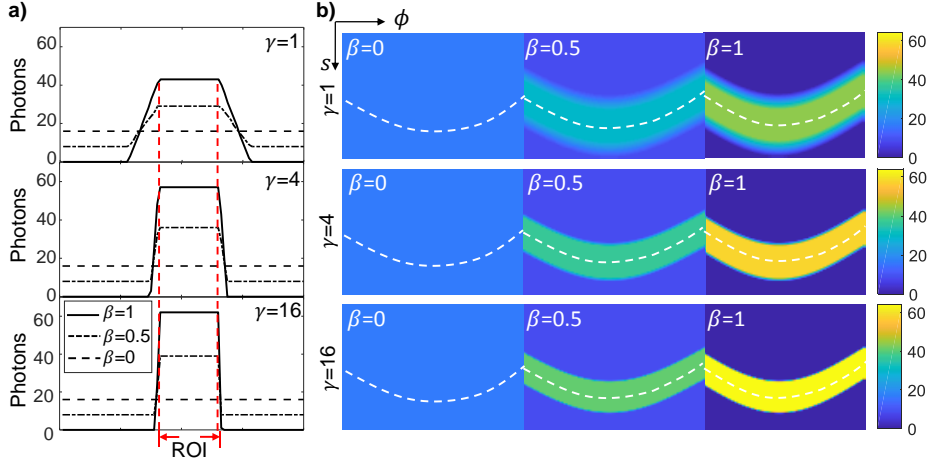

Fig. 2: Photon distribution profiles used in the simulation. (a) Photon allocation profiles at $\gamma=1$, 4 and 16 under $0°$ projection. (b) two-dimensional photon allocation maps for $\beta=0$, 0.5 and 1, $\gamma=1$, 4 and 16.

where $\delta x$ is the translation step size; $\sigma$ is the radius of ROI; $\beta$ controls the interior/exterior ratio; $\Delta$ denotes the width of the transition region where the photon number drops from maximum to the minimum; when $\Delta=0$, the photon allocation profile becomes a rectangular shape. In our simulation, we define $\gamma = \sigma/\Delta$ as the shape parameter that describes the normalized slope of the photon numbers across the ROI boundary. $s_c$ is the center coordinate of ROI at different projections, and is calculated via

$$
s_c = s_{offset} \sin(\phi + \phi_{offset}) \tag{23}
$$

where $s_{offset}$ is the distance between the ROI center and the rotation center; $\phi_{offset}$ is the azimuthal coordinate of the ROI center with respect to the rotation center. Fig. 2 (a) plots the photon allocation profile along the translation direction, $s$, with $\gamma=1$, 4 and 16 under $0°$ projection. For phantom 1, the two-dimensional phantom allocation maps for $\beta=0$, 0.5 and 1 are shown in Fig. 2 (b). The white, dashed curves mark the trajectory of ROI center as the sample undergoes $180°$ rotation. The reconstruction quality is assessed with the normalized MSE (NMSE) to facilitate the comparison among different ROIs and samples.

$$
NMSE = MSE / \sum_{i_x, i_y=1}^{n_x, n_y} \|\mathbf{w}^T \mathbf{f}\|^2 \tag{24}
$$

Our simulation goal was to find the best photon allocation strategy, expressed by the two parameters $(\beta, \gamma)$, that minimizes the NMSE between the reconstruction and the object within ROI.

### B. Simulation setup

The 2D photon map $\mathbf{r}$ was calculated from Eq. (22) and rounded to the nearest integer. Each pencil-beam measurement $g_j$ was simulated by summing up $r_j$ geometric random numbers. In case zero photons were assigned to a particular pencil beam, the corresponding row that represents this measurement was removed from the CT matrix $\mathbf{A}$. In total, 11 interior/exterior ratios $\beta$ (from 0 to 1 at 0.1 step) and 3 shape parameters $\gamma=1$, 4, and 16 were combined as different photon allocation strategies in our simulations. For each photon map $\mathbf{r}$, we ran 15 instances of simulated measurements $\mathbf{g}$ and reconstructions $\hat{\mathbf{f}}$. Based on the 15 reconstruction instances $\hat{\mathbf{f}}$, we calculated the NMSE within the ROI, and plotted its average and standard deviation as the error bar. We also recorded the pixel-wise mean, variance and MSE map for comparison with our analytical approximation.

### C. Experiment setup

The experimental photon-counting system was implemented by connecting the electrically censored pulses from the Si-PIN detector (X-123, AMPTEK) to a data acquisition device (USB6353, National Instrument) operating in the edge-counting mode. The detector system was configured to run at a counting interval of $\Delta t=10\mu s$. The X-ray source was a copper-anode tube (XRT60, Proto Manufacturing) operating at 12kV. The current of the tube had been reduced to 1mA so that no two photons could overlap and register in the same time interval. The incident X-ray beam was collimated by a pair of 0.5mm pinholes located at 20cm away from the X-ray focus. With this setup geometry, the probability of detecting one photon within one $\Delta t$ time interval from the collimated beam was $\lambda$=0.015. The sample was mounted on a rotational stage (RV120PP, Newport) and a linear stage (UTM150CC, Newport) for pencil-beam CT scan.

An acrylic resolution target was scanned in our experiments. The target was laser-machined with 0.5-1.0mm line-widths groups. The 0.6mm group was defined as the ROI. The sample was translated at a step size of 0.2mm for 81 steps in total, and rotated by 2° steps to cover 180° projections. For each pencil-beam measurement, we collected the entire 1s time stamps and used all the detected photons to form a reference image. The low-photon measurement had a detected photon budget of 16 photons per pencil beam on average. We extracted the first $r_j$ photons from the 1s time stamp according to our optimized strategy for reconstruction.

### D. Reconstruction algorithm

The optimization problem in Eq. (6) is solved numerically with a modified SPIRAL-TAP[29], which is a gradient-descent algorithm combined with regularization in each iteration. The gradient and Hessian of our negative binomial log-likelihood are respectively

$$\nabla l(\mathbf{f}) = \mathbf{A}^T \left( \mathbf{r} - \frac{\lambda(\mathbf{g} - \mathbf{r}) \odot \exp(-\mathbf{Af})}{1 - \lambda \exp(-\mathbf{Af})} \right), \quad (25)$$

$$\mathbf{H}(l(\mathbf{f})) = \mathbf{A}^T \left( \frac{\lambda(\mathbf{g} - \mathbf{r}) \odot \exp(-\mathbf{Af})}{(1 - \lambda \exp(-\mathbf{Af}))^2} \right) \mathbf{A}. \quad (26)$$

Different from the Hessian matrix in the original SPIRAL-TAP algorithm, Eq. (26) contains a singular point on the denominator, which indicates that the Hessian matrix is positive-definite only on the non-negative domain $\mathbf{f} \in \mathbb{R}_+^n$. To avoid moving the solution across the singular point of the Hessian matrix, we enforced a minimum value of $\alpha$, $\alpha_{min}$, in each iteration if the accepted $\alpha$ value according to Ref. [29] is smaller than $\alpha_{min}$.

The regularization step in our modified SPIRAL-TAP implements either TV or L2-norm constraint. The TV regularization minimizes the total-variance semi-norm

$$u(\mathbf{f}) = \tau \sum_{i_x=1}^{n_x} \sum_{i_y=1}^{n_y} \sqrt{(f_{i_y+1, i_x} - f_{i_y, i_x})^2 + (f_{i_y, i_x+1} - f_{i_y, i_x})^2} \quad (27)$$

and enforces non-negativity with the FISTA algorithm [30]. The L2-norm constraint belongs to a special case in the Gaussian prior family, with $\mathbf{D}$ chosen as the identity matrix $\mathbf{I}_n$ in Eq. (12). This leads to the solution to the regularization step at $k$-th iteration in SPIRAL-TAP

$$\mathbf{f}^{k+1} = \{\max\left(0, \frac{f_i^{k+1, temp}}{1 + \tau/\alpha^k}\right), i = 1, 2, ..., n\} \quad (28)$$

where $1/\alpha^k$ is the step size along the gradient; $\mathbf{f}^{k+1, temp}$ is the intermediate result after the $k$-th gradient-descent step; and $\mathbf{f}^{k+1}$ is the final result after regularization. The negative values in $\mathbf{f}^{k+1, temp}$ are replaced with 0 to enforce non-negativity. For both constraints, the regularization strength, $\tau$, ranging from $10^1$ to $10^3$ at $10^{0.5}$ step, was optimized for minimum NMSE in the reconstruction.

## IV. NUMERICAL AND EXPERIMENTAL RESULTS

### A. Phantom simulations with L2-norm constraint

#### 1) The effect of $\gamma$ on NMSE

We first evaluated the ROI reconstruction performances with different photon allocation strategies using the abdomen phantom. The regularization parameter that yielded minimal NMSE for each photon map was selected. The total detected photon budget was $1.15 \times 10^6$, corresponding to 16 photon counts per pencil beam on average. Fig. 3 plots the reconstruction NMSE within the ROI from our analytical predictions (Eq. (21)) and numerical experiments in terms of interior/exterior ratio $\beta$ and shape parameter $\gamma$. As $\beta$ increases from 0 up to 0.9, more photons are allocated to the ROI, therefore decreasing its reconstruction NMSE. This simulation trend was correctly predicted in our analytical model, with 86% of the predictions lying within the standard deviation of the simulation NMSE. The optimal $\beta$ and the minimum NMSE are summarized in Table 1 for $\gamma$=1, 4 and 16. The best photon strategy is $\beta$=0.8, $\gamma$=16 in the numerical simulation. It is worth noting that for $\beta$<0.9, the NMSE of $\gamma$=1 is generally higher than $\gamma$=4 and 16. This is because for the same $\beta$, $\gamma$=1 allocates more photons in the vicinity outside the ROI boundary, which reduces the ROI photon counts as a result of the fixed total photon budget. Therefore a shape parameter $\gamma$>1 is generally preferred.

TABLE 1:
THE BEST $\beta$ AND NMSES FOR DIFFERENT SHAPE PARAMETER $\gamma$

| Shape parameter $\gamma$ | $\gamma$=1 | $\gamma$=4 | $\gamma$=16 |
|---|---|---|---|
| Optimal $\beta$ | 0.9 | 0.7 | 0.8 |
| NMSE | 0.52±0.04% | 0.53±0.02% | 0.51±0.02% |

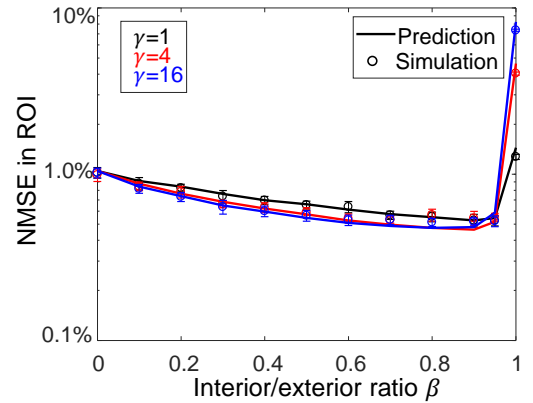

Fig. 3: Simulated and predicted ROI reconstruction NMSE vs. the interior/exterior ratio $\beta$ for $\gamma$=1, 4, and 16.

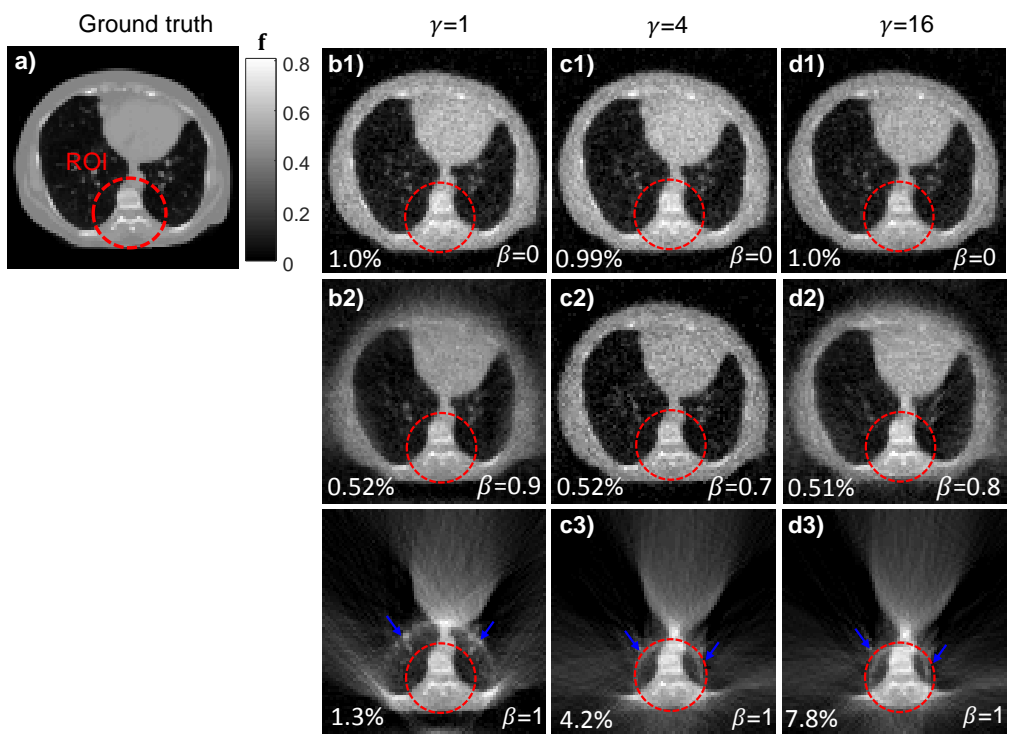

Fig. 4: (a) Ground truth of the abdomen phantom used in the simulation. (b-d) Reconstruction of the abdomen phantom with different shape parameters $\gamma$ for (b) $\beta$=0 (uniform), (c) optimized $\beta$, and (d) $\beta$=1. The ROI is marked by the red circle. The blue arrows highlight the truncation artifacts that appear at $\beta$=1. The numbers in the left bottom of each reconstruction indicate the NMSE within ROI.

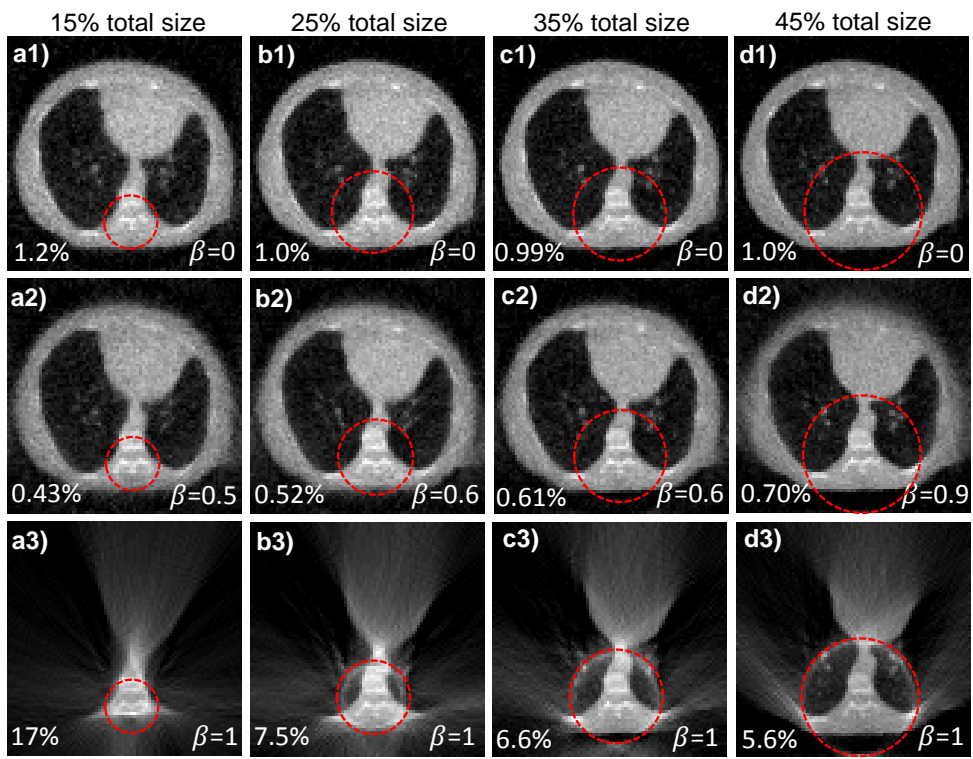

Fig. 5: Reconstruction of the abdomen phantom with ROI sizes equal to (a) 15% (b) 25% (c) 35% (d) 45% of the total size, for (1) $\beta$=0 (uniform), (2) optimized $\beta$, and (3) $\beta$=1 and $\gamma$=16. The ROI is marked by the red circle. The numbers in the left bottom of each reconstruction indicate the NMSE within ROI.

The increase in NMSE when $\beta$ approaches 1 is primarily attributed to the absence of the exterior projection measurement. Fig. 4 exemplifies several reconstruction instances at different $(\beta, \gamma)$ combinations. The ground truth and ROI of the abdomen phantom are shown in Fig. 4 (a) for comparison. Fig. 4 (b-d) show the reconstruction at $\beta$=0, the optimized $\beta$ in the simulation, and $\beta$=1. At $\beta$=1, some regions beyond the ROI boundary received no photons in the simulation, and thus were excluded from the measurement. This creates the visible artifacts around the boundary between measured and unmeasured regions in the reconstruction (Fig. 4 (b3-d3)), which are consistent with the truncation artifacts associated with the non-localized filtered back-projection (FBP) kernel [15], [31]. The photon allocation for $\beta$=1, $\gamma$=16 is equivalent to the projection truncation in interior tomography, in which no projections outside the ROI is measured. The truncation artifacts contribute to a large reconstruction error (7.8% NMSE in ROI), especially on the ROI boundary. Our simulation and analytical predictions suggest that the best strategy ($\beta$=0.8, $\gamma$=16) is to deposit 20% of the total photon budget outside ROI, which reduces the NMSE in ROI by ~15 times compared to concentrating all available photon budgets in the ROI ($\beta$=1, $\gamma$=16).

### 2) The effect of ROI size on NMSE

As $\beta$ increases from 0 to the optimal value, the general trend of NMSE is to decrease, then increase dramatically as $\beta$ passes the optimum and missing measurements start to appear. Since this NMSE trend holds for different $\gamma$s, we focus on the case of $\gamma$=16, which is the optimal choice in Sec. 4.A1, in the following discussions. Fig. 6 plots the NMSE vs. $\beta$ with different ROI diameters $2\sigma$, normalized with respect to the total photon size. For this simulation, we scaled up/down the total photon budget $I_0$ so that for the same $\beta$, the average number of photons per beam dedicated to the ROI, $I_0\beta/(2\sigma m_s)$, remained unchanged regardless of the ROI size. All the ROIs in this simulation are centered on the same location as in Fig. 4 (a). Fig. 5 shows the reconstruction instances for $\beta$=0, optimized $\beta$, and $\beta$=1 when ROI diameter equals 15%, 25%, 35% and 45% of the full phantom dimension, respectively. The optimal $\beta$ and NMSE are summarized in Table 2. These results indicate that the optimized $\beta$ parameter is mainly determined by the portion of ROI within the whole sample. A smaller ROI generally requires lower $\beta$ for optimal reconstruction performance.

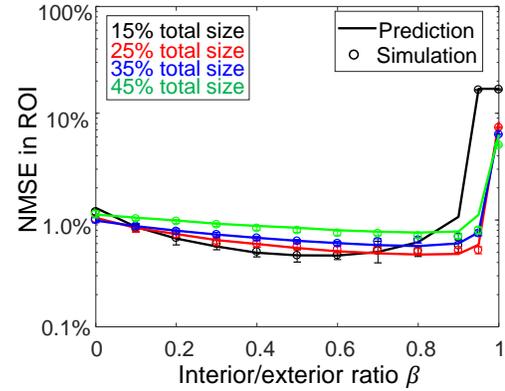

Fig. 6. ROI reconstruction NMSE vs. the interior/exterior ratio $\beta$ for ROI diameters equal to 15%, 25%, 35% and 45% of the phantom size, with $\gamma$=16.

### 3) The effect of total photon budget on NMSE

In the previous examples, we have shown that an interior photon allocation scheme, $\beta$=1, is generally not favorable for ROI reconstruction. Here, we focus on the allocation strategies for $\beta$<1 at different total photon budgets, ranging from an average of 16 photons/beam to 1024 photons/beam. Fig. 7 plots the NMSE as a function of $\beta$ with $\gamma$=16 at different photon budgets for the ROI shown in Fig. 4 (a). The NMSE of uniform and optimized photon allocations are summarized in Table 3. Fig. 8 shows the phantom reconstruction instances with respect to different photon budgets and $\beta$. As the total photon budget increases, the reconstruction quality improves for all $\beta$ values; the relative difference in the ROI reconstruction NMSE between uniform and optimized strategies decreases from ~50% down to 35%. These results suggest that the improvement in the reconstruction performance of a carefully optimized photon allocation strategy is especially prominent for low photon-budget scenarios.

TABLE 3:
THE BEST $\beta$ AND RELATIVE CHANGES IN NMSES BETWEEN UNIFORM AND OPTIMIZED PHOTON ALLOCATION WITH DIFFERENT TOTAL PHOTON BUDGETS

| Average photons per beam | 16 | 64 | 256 | 1024 |
|---|---|---|---|---|
| Optimal $\beta$ | 0.9 | 0.8 | 0.7 | 0.7 |
| Optimal distribution NMSE | 1.0% ±0.07% | 0.39% ±0.02% | 0.19% ±0.01% | 0.12% ±0.01% |
| Uniform distribution NMSE | 0.51% ±0.03% | 0.24% ±0.02% | 0.12% ±0.01% | 0.078% ±0.005% |
| Relative change in NMSE | 49% | 38% | 36% | 35% |

TABLE 2:
THE BEST $\beta$ AND NMSES FOR DIFFERENT ROI SIZES

| ROI size ratio | 15% | 25% | 35% | 45% |
|---|---|---|---|---|
| Optimal $\beta$ | 0.5 | 0.6 | 0.6 | 0.9 |
| NMSE | 0.46% ±0.06% | 0.52% ±0.03% | 0.60% ±0.03% | 0.70% ±0.03% |

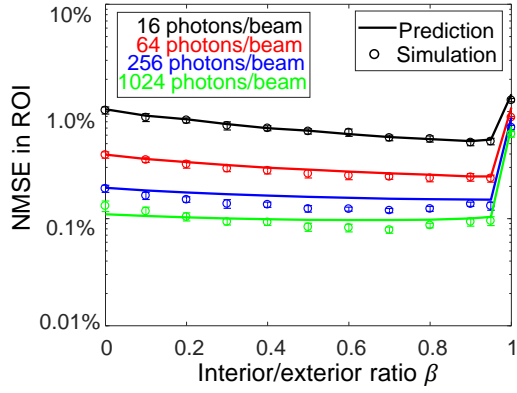

Fig. 7: ROI reconstruction NMSE vs. the interior/exterior ratio $\beta$ at different total photon budget, with $\gamma$=16. The black, red, blue and green curves represent 16, 64, 256 and 1024 detected photons per beam on average.

### B. Phantom simulation with total-variance (TV) prior

In this section, we demonstrate the photon allocation strategies applied to ROI reconstructions with TV prior. The object in this simulation is a piecewise-constant Shepp-Logan phantom (Fig. 9 (a)). Fig. 9 (b) plots the NMSE of the ROI reconstruction with different ($\beta,\gamma$) combinations compared to the ground truth. Fig. 9 (c, d) shows the photon map and ROI reconstruction at the optimized strategy and pure interior measurement. The best reconstruction performance was attained at $\beta$=0.7, $\gamma$=4, with an NMSE of 0.40%±0.04%. The optimal regularization parameter $\tau$ was $10^{1.5}$. In contrast, the reconstruction from truncated projection ($\beta$=1, $\gamma$=16) had an

NMSE of 4.6%±0.2%, which was obtained at the optimal $\tau$=$10^{2.5}$. We speculate that the smooth and low-contrast ROI reconstruction in Fig. 9 (c2) is mainly due to the large regularization parameter used in ROI reconstruction. This comparison demonstrates that allocating some photons to the exterior region obviates the need for a strong piecewise constant regularizer, thus achieving superior reconstruction quality than TV-based interior reconstruction.

### C. Resolution target imaging

Based on the high consistency between the analytical model and the numerical estimator with L2 prior, we were able to approximately predict the optimized strategy to allocate a fixed photon budget for an experimental object. Fig. 10 shows a full-scan CT image of an acrylic resolution target acquired with 1s integration time per pencil beam, which was used as the ground truth for reference. The average photon count per beam was 589 within 1s. The 0.6mm line-width group within the target was defined as the ROI. The ROI reconstruction was performed at 16 photons per beam on average. Fig. 10 (b) plots the analytically-predicted reconstruction MSE (compared to the reference in (a)) inside the ROI as a function of interior/exterior ratio $\beta$ and the shape parameter $\gamma$. For each photon allocation strategy, we calculated the bias and variance with 6 regularization parameters $\tau$ ranging from $10^{1}$~$10^{3.5}$, and selected the $\tau$ that yields the minimal predicted MSE for use in the SPIRAL reconstruction. From our prediction, the smallest ROI reconstruction error was attained at $\beta$=0.7, $\gamma$=4 with the regularization parameter $\tau$=$10^{2}$. Fig. 10 (d, e) shows the measured time intervals **g** (d1-d3) and reconstructions (e1-e3) from 3 photon allocation maps **r** (c1-c3), which correspond to uniform (c1), optimized (c2) and ROI-only (c3) strategies, respectively. The optimized strategy had a reconstruction NMSE of 2.8% at ~30 times reduced photon budget compared to the reference image.

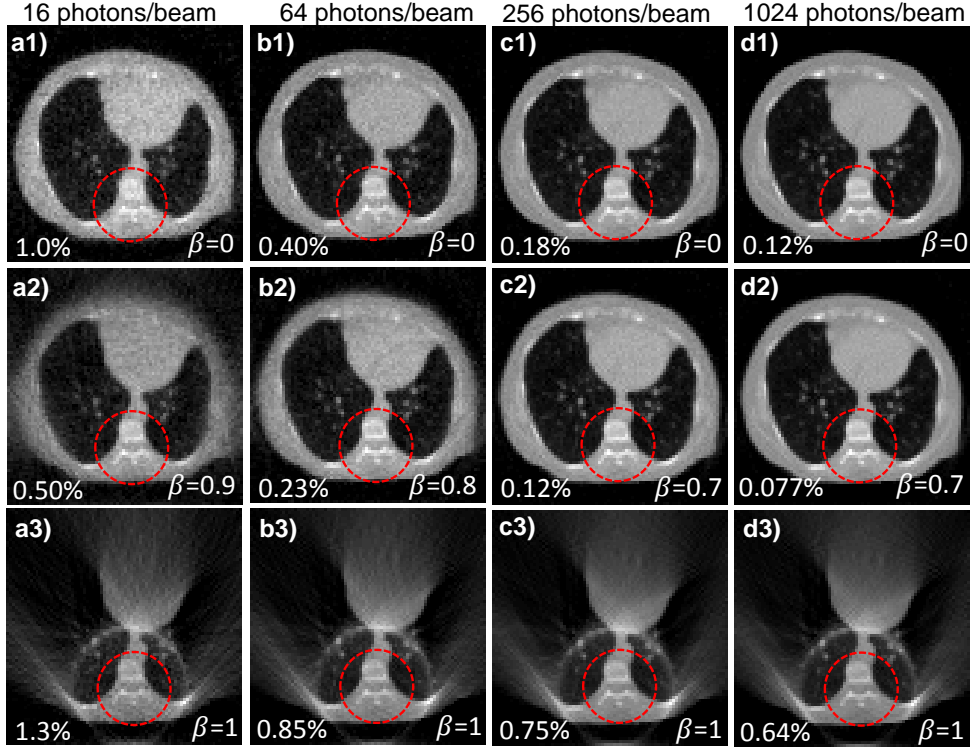

Fig. 8: Reconstruction of the abdomen phantom with different total photon budget (a) 16, (b) 64, (c) 256 and (d) 1024 photons per beam on average for (1) $\beta$=0 (uniform distribution), (2) optimized $\beta$, and (3) $\beta$=1 with $\gamma$=16. The ROI is marked by the red circle. The numbers in the left bottom of each reconstruction indicate the NMSE within ROI.

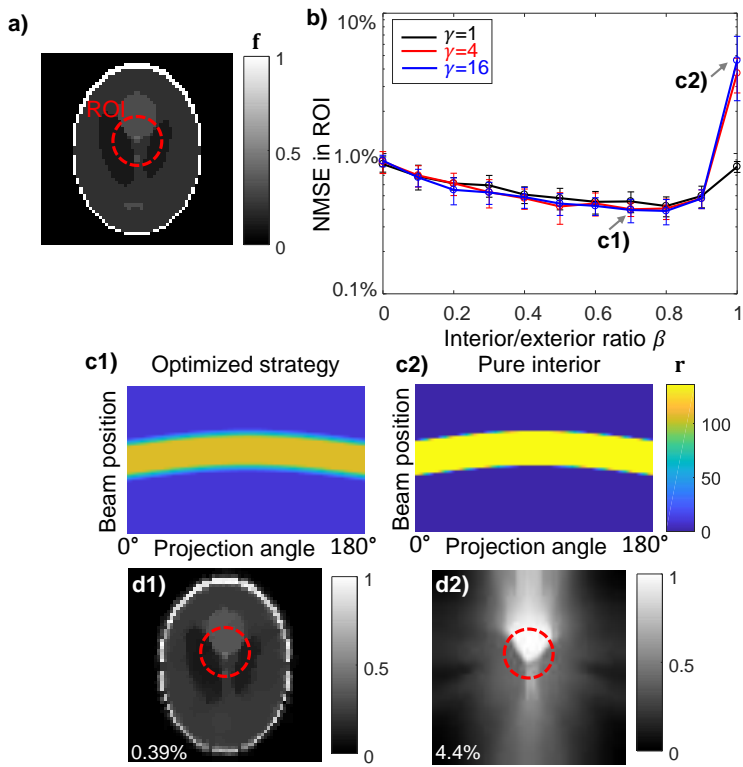

Fig. 9: Photon allocation strategy for TV prior. (a) Shepp-Logan phantom and ROI in the simulation. (b) NMSE in ROI vs. interior/exterior ratio $\beta$ for $\gamma=1$, 4, and 16. (c) Photon allocation map for (c1) optimized strategy ($\beta=0.7$, $\gamma=4$) and (c2) conventional truncated projection ($\beta=1$, $\gamma=16$). (d1, d2) Reconstruction from (c1), (c2), respectively. The ROI is highlighted in (d1, d2). The numbers in the left bottom of each reconstruction indicate the NMSE within ROI.

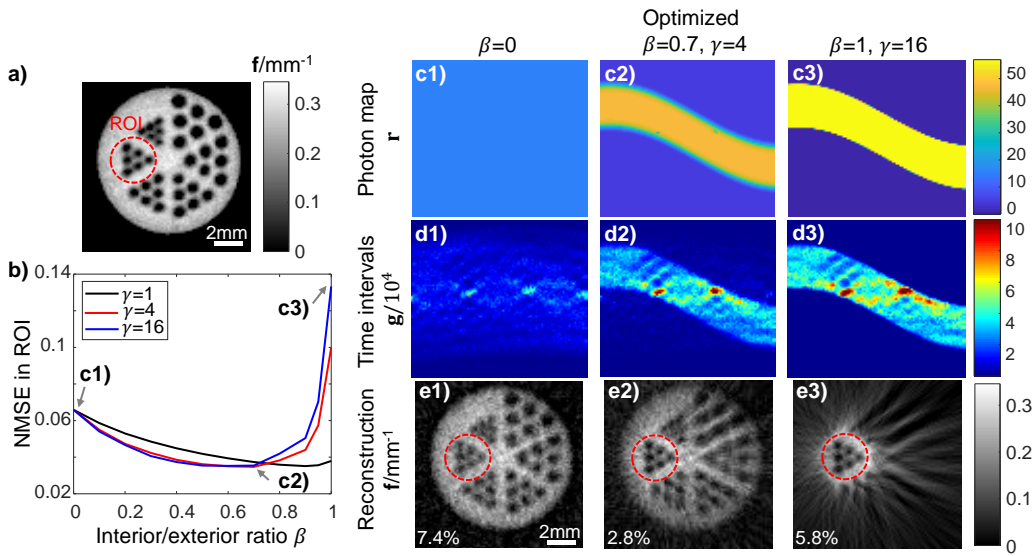

Fig. 10: Comparison between different photon allocation strategies for the resolution target. (a) Full-scan CT image of the resolution target. The ROI covers the 0.6mm line-width group. (b) Predicted reconstruction MSE in ROI with respect to different photon allocation strategies, expressed in terms of the interior/exterior ratio $\beta$ and the shape parameter $\gamma$. (c–e) Examples of the photon allocation strategies, experimental measurements and the corresponding ROI reconstructions from (1) uniform photon counts (2) optimized photon allocation map (3) interior measurement with $\beta=1$, $\gamma=16$. The average photon count was 16 per beam. All scale bars represent 2mm. The ROI is marked by the red, dashed circle in (a) and (e). The numbers in the left bottom of each reconstruction indicate the MSE within ROI.

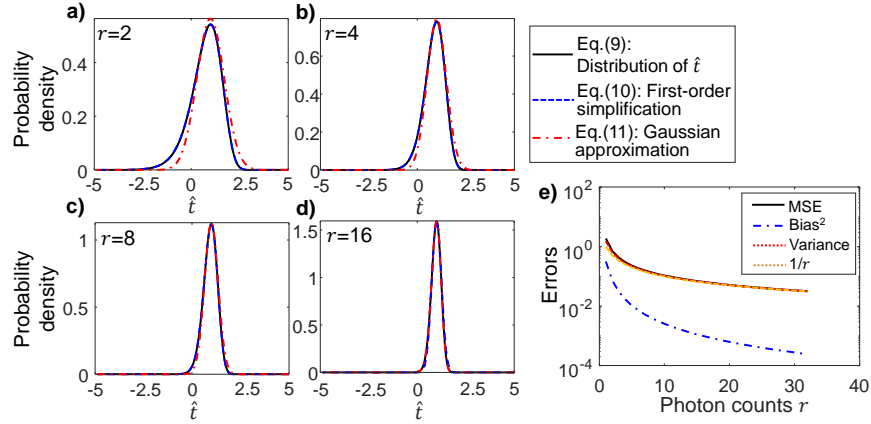

Fig. 11: Comparison between the distribution of $\hat{t}$ and our approximations. (a-d) Original distribution of $\hat{t}$, first-order simplification and Gaussian approximation for $r$=2, 4, 8 and 16, $t$=1.0. (e) Bias², variance and MSE of $\hat{t}$ with respect to the ground truth $t$. The dashed, orange curve plots the variance of Gaussian-approximated, unbiased distribution of $\hat{t}$ (Eq. (11)).

## V. DISCUSSIONS

### A. Approximations on the distribution of $\hat{t}$

In Section 3C, we approximate the distribution of the measured CT line integral $\hat{t}$ with a Gaussian distribution. Fig. 11 compares the distribution $q_j(\hat{t}_j|t_j;r_j)$(Eq. (9)), the first-order simplification (Eq. (10)) and our Gaussian approximation (Eq. (11)) without remainder $R(\hat{t}_j)$ for $r_j$ ranging from 1 to 32, and $t_j$=1.0. The difference between $q_j(\hat{t}_j|t_j;r_j)$ and the first-order simplification is negligible. The discrepancy mainly arises from the remainder term $R(\hat{t}_j)$ in Eq. (11), which introduces negative skew on the normal distribution and lowers the mean of $\hat{t}_j$. Fig. 11 (e) quantifies the bias, variance and MSE of the distribution $q_j(\hat{t}_j|t_j;r_j)$ with respect to the ground truth $t_j$. The variance and MSE of the unbiased Gaussian approximation, $1/r_j$, is also plotted for comparison. As the photon number increases to 16 and above, the bias contributes to less than 1% in the MSE, which indicates that the unbiased normal distribution is a good approximation to $q_j(\hat{t}_j|t_j;r_j)$. Notice that for a different ground truth $t_j$, Eq. (10) implies that the peak of the distributions in Fig. 11 (a-d) would shift, while their shapes remain the same.

### B. Approximations on the MAP objective function

In our analytical models and numerical simulations, the regularization parameter was selected by minimizing the reconstruction MSE. Fig. 12 (a) plots the bias, variance and MSE of both the negative binomial (Eq. (6)) and the approximated (Eq. (15)) estimators with respect to different regularization levels $\tau$. Here we chose as an example the optimized photon allocation strategy ($\beta$=0.8, $\gamma$=16) of the abdomen phantom in Fig. 12. All plotted values are normalized with respect to the L2-norm of the object ground truth in ROI. The bias-variance trade-off in the MSE of the estimator can be clearly seen in both analytical and modified SPIRAL estimator. Both curves predict a choice of $\tau$=10² to minimize the reconstruction MSE.

Fig. 12 (b-c) details the pixel-wise bias square, variance and MSE map from the analytical approximation and modified SPIRAL estimator at $\tau$=10². Due to the higher photon count, the variance within ROI is smaller than the exterior region in both (b2) and (c2). The variance map of (c2) shows object-dependent features that do not exist in the analytical approximation (Eq. (20)). This is because the modified

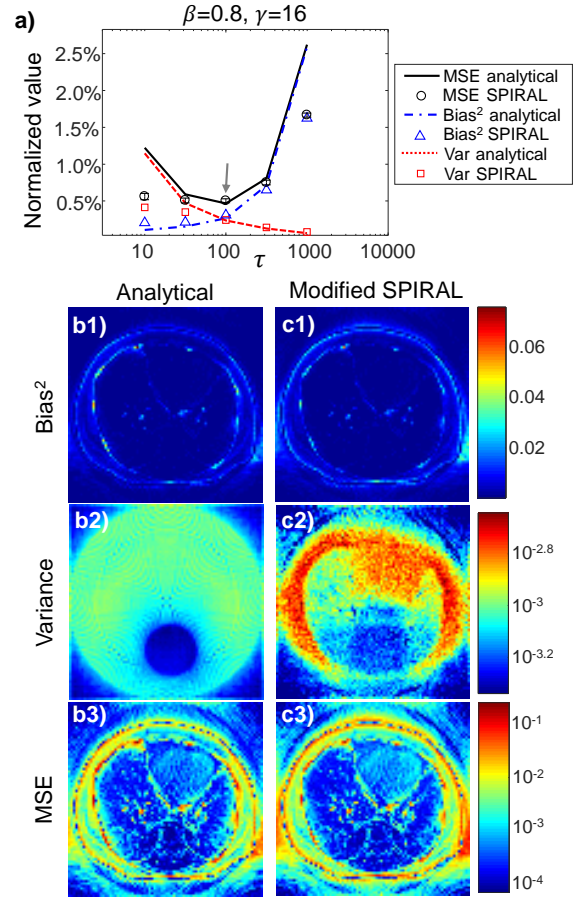

Fig. 12: Comparison between the predicted and simulated bias, variance and MSE map, normalized with respect to the L2-norm of object ground truth in ROI. (a) Bias square, variance and MSE within the ROI with respect to the regularization parameter $\tau$ for the photon map $\beta$=0.8, $\gamma$=16. (b-c) Predicted and simulated (1) mean, (2) variance and (3) MSE map of the object from (b) analytical approximation, (c) SPIRAL with negative binomial objective function.

SPIRAL estimator with positivity constraint (Eq. (6)) is a non-linear function of $\hat{t}$. Table 4 summarizes the NMSE within the ROI and whole sample. Despite the discrepancy in the variance maps between modified SPIRAL and our approximation, for $\tau$ on the order of $\sim 10^2$, which we pick in the reconstruction, the NMSE of the numerical simulation matches well with Eq. (21) within the ROI.

TABLE 4:
COMPARISON OF NMSE AMONG THE ANALYTICAL
APPROXIMATION AND MODIFIED SPIRAL WITH NEGATIVE
BINOMIAL LIKELIHOOD

| NMSE | Analytical approximation | Modified-SPIRAL |
|---|---|---|
| Within ROI | 0.48% | 0.51% |
| Whole phantom | 5.8% | 7.3% |

## VI. Conclusion

In summary, we have proposed a few-photon measurement framework and its corresponding statistical reconstruction algorithm optimized for region-of-interest x-ray computed tomography. The demonstrated framework is capable of reconstructing an ROI with an average detected photon budget of ~16 photons per beam. We model the numerical reconstruction as a Bayesian estimator, and study its bias, variance and MSE with various levels of regularization. The analytical model with L2 prior shows high consistency with the simulation, and correctly predicts the trend of MSE for different photon allocation maps. Although predicting MSE involves the ground truth of the attenuation map **f**, this information can be obtained from a fast, pre-diagnostic scan, which has already been practiced in a number of multi-resolution, region-of-interest CT systems [13], [22], [32].

The combination of negative binomial photon statistics with an optimized photon allocation strategy presents a novel approach to efficiently utilizing the available photon budget, which is especially attractive for low-photon scenarios. By optimizing the two parameters controlling the profile of the photon allocation, we have demonstrated ~2-fold improvement in ROI reconstruction compared to uniformly allocating the same photon budget throughout the sample, and 10~15-fold improvement compared to concentrating all available photons in the ROI. In our numerical experiments, we have discovered that the optimized photon ratio between ROI and exterior region is primarily determined by the ROI diameter relative to the whole object size. A smaller ROI requires more photons outside ROI to lower the NMSE contributed from the noisy exterior region. In a real CT experiment, we were able to faithfully reconstruct an ROI of the resolution target at ~30 times reduced total photon budget.

The proposed time-stamp photon-counting interior tomography scheme can be extended beyond the demonstrated pencil-beam CT system. The energy discrimination capability of the PCDs allows simultaneous acquisition of attenuation maps in different x-ray energy (wavelength) channels, which facilitates the integration of our photon-counting framework into existing dual-energy or multi-energy CT modalities with a PCD array for parallelization. The high photon-efficiency associated with the x-ray PCDs could extend our model to the reconstruction of x-ray diffraction tomography [33], where the diffraction signal is intrinsically ~3 orders of magnitude lower than the transmission signal. We envision that our photon-counting tomography framework could be applied to photon-starved environments for both x-ray and visible imaging systems.

## Acknowledgment

The authors would like to thank Dr. Alexander Katsevich (Department of Mathematics, University of Central Florida) for helpful discussions.